\documentclass[traditabstract]{aa}

\usepackage{amssymb}
\usepackage[usenames, dvipsnames]{color}
\usepackage{graphicx}
\usepackage{natbib}
\usepackage[rightcaption]{sidecap}
\usepackage{txfonts}
\usepackage{units}
\usepackage{url}
\usepackage{pifont}
\usepackage[modulo, switch]{lineno}
\usepackage{hyperref}
\hypersetup{
     colorlinks   = true,
     citecolor    = blue,
     urlcolor     = blue,
     linkcolor    = blue,
     anchorcolor  = blue
}
        
\definecolor{blue}{rgb}{0.00, 0.00, 1.00}
\definecolor{red}{rgb}{0.86, 0.08, 0.24}
\definecolor{orange}{rgb}{1.00, 0.55, 0.00}
\definecolor{darkblue}{rgb}{0.00, 0.00, 0.55}
\definecolor{green}{rgb}{0.00, 0.39, 0.00}
\definecolor{pink}{rgb}{1.000000,0.078431,0.576471}

\definecolor{myDarkRed}{rgb}{0.698, 0.094, 0.133}

\usepackage{lipsum}

\begin{document} 


\title{Capabilities of bisector analysis of the \ion{Si}{i} 10827~\AA \ line for estimating line-of-sight velocities in the quiet Sun}
\titlerunning{On the capabilities of bisector analysis of the \ion{Si}{i} 10827~\AA}

\author{%
    S.J. Gonz{\'a}lez Manrique\inst{1},
    C. Quintero Noda\inst{2,3},
    C. Kuckein\inst{4},
    B. Ruiz Cobo\inst{5,6},  \and
    M. Carlsson\inst{2,3}
    }
   
\institute{%
    $^1$ Astronomical Institute, Slovak Academy of Sciences, 
         05960 Tatransk\'{a} Lomnica, Slovak Republic\\
    $^2$ Rosseland Centre for Solar Physics, University of Oslo, 
         P.O. Box 1029 Blindern, N-0315 Oslo, Norway\\
    $^3$ Institute of Theoretical Astrophysics, University of Oslo, 
         P.O. Box 1029 Blindern, N-0315 Oslo, Norway\\
    $^4$ Leibniz-Institut f{\"u}r Astrophysik Potsdam (AIP),
         An der Sternwarte 16, 
         14482 Potsdam, Germany\\
    $^5$ Instituto de Astrof\'{i}sica de Canarias, 
         V\'{i}a L\'{a}ctea s/n, 38205 La Laguna, Tenerife, Spain\\
    $^6$ Departamento de Astrof\'{i}sica, Universidad de La Laguna,  
         38206, La Laguna, Tenerife, Spain\\
    \email{smanrique@ta3.sk}}

\date{Received \today; accepted later}

\abstract
{We examine the capabilities of a fast and simple method to infer line-of-sight (LOS) velocities from observations of the photospheric \ion{Si}{i} 10827~\AA\ line. This spectral line is routinely observed together with the chromospheric \ion{He}{i} 10830~\AA \ triplet as it helps to constrain the atmospheric parameters. We study the accuracy of bisector analysis and a line core fit of \ion{Si}{i} 10827~\AA. We employ synthetic profiles starting from the Bifrost enhanced network simulation. The profiles are computed solving the radiative transfer equation, including non-local thermodynamic equilibrium effects on the determination of the atomic level populations of \ion{Si}{i}. We found a good correlation between the inferred velocities from bisectors taken at different line profile intensities and the original simulation velocity at given optical depths. This good correlation means that we can associate bisectors taken at different line-profile percentages with atmospheric layers that linearly increase as we scan lower spectral line intensities. We also determined that a fit to the line-core intensity is robust and reliable, providing information about atmospheric layers that are above those accessible through bisectors. Therefore, by combining both methods on the \ion{Si}{i} 10827~\AA\ line, we can seamlessly trace the quiet-Sun LOS velocity stratification from the deep photosphere to higher layers until around $\log \tau = -3.5$ in a fast and straightforward way. This method is ideal for generating quick-look reference images for future missions like the Daniel K. Inoue Solar Telescope and the European Solar Telescope, for example.}

    
\authorrunning{Gonz{\'a}lez Manrique et al.}

\maketitle
   

\section{Introduction}\label{SEC1}

The next decade promises breakthroughs in solar ground-based observations. Two of the next generation of solar observatories come in the form of the Daniel K. Inoue Solar Telescope \citep[DKIST,][]{Keil2011} and the European Solar Telescope \citep[EST,][]{Collados2013}, both with four-metre apertures. The former will start operations in 2020 and the latter is expected to be finalised in 2028. Both telescopes aim to achieve high spatial resolution and signal-to-noise-ratio observations. Also, they will perform spectro-polarimetric measurements of multiple spectral lines whose formation spans from the lower photosphere and beyond. There is a new generation of inversion codes, such as for example NICOLE \citep{SocasNavarro2015}, SNAPI \citep{Milic2018}, STiC \citep{delaCruzRodriguez2019}, or FIRTEZ-dz \citep{PastorYabar2019}, that aim to infer the physical information of the atmospheric parameters by fitting multiple spectral lines at the same time. However, there is also a need for the development and testing of more simple and less computationally demanding techniques that allow the extraction of basic information; for example, quick-look visualisation of the observed data. Demand for such techniques stems from the large amount of data that will be generated by the above telescopes each day, an amount of data that is too large to be inverted in reasonable times with the above-mentioned codes.

\begin{figure}
\begin{center} 
 \includegraphics[trim=0 0 0 0,width=8.5cm]{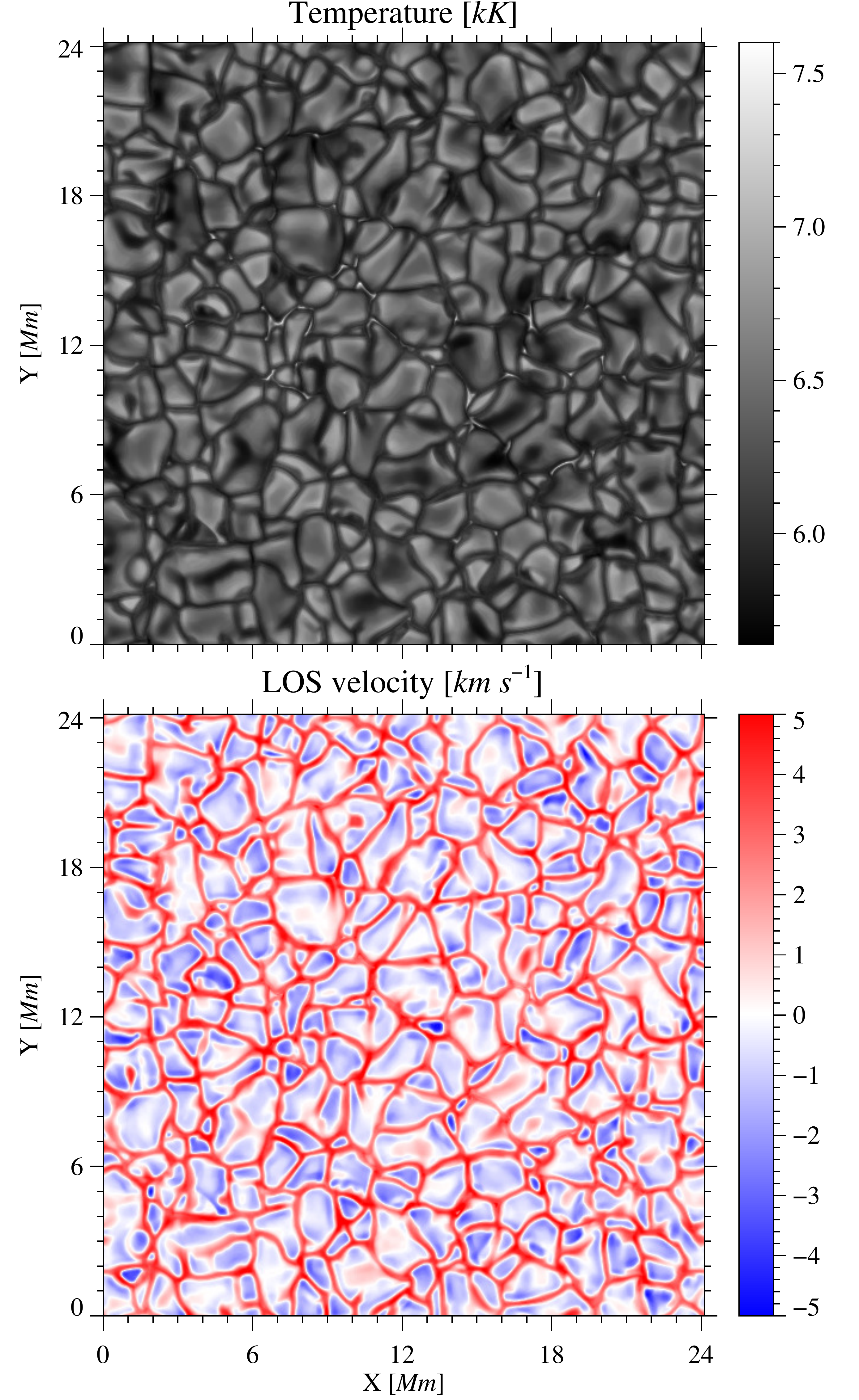}
 \vspace{-0.2cm}
 \caption{Snapshot 385 from the {\sc bifrost} enhanced network simulation. 
          The top panel shows the gas temperature while the bottom panel displays the LOS 
          velocity (positive means downflow) at 
          the optical depth $\log \tau_{5000}=0$.}
 \label{context}
 \end{center}
\end{figure}

We can complement those advanced techniques with simple methods like the weak-field approximation \citep[e.g.][]{Landi2004} for inferring the magnetic field of DKIST and EST target spectral lines \citep[e.g.][]{Centeno2018,Shchukina2019} or more elaborate techniques like Milne-Eddington (ME) inversions \citep[][]{Landolfi1982}. The latter technique has been routinely used on spectropolarimetric data from space missions, for example using the MERLIN \citep{Lites2007} and VFISV \citep{Borrero2011} codes. Milne-Eddington inversions can provide the three components of the magnetic field and the line-of-sight (LOS) velocity  in a short
time. These codes infer this information as a single value, which is usually associated with the height where the fitted line is most sensitive to that atmospheric parameter \citep[][]{delToroIniesta2016}. However, there are additional tools that can be used as fast approaches, and they simultaneously provide information about the vertical stratification of the atmospheric parameters.

Bisector methods proposed by \citet[][]{Kulander1966}, which are also as fast as any ME approach, can derive the LOS velocity at different atmospheric layers. These methods were designed based on the fact that spectral lines often show asymmetries in their profiles, which are usually associated with velocity gradients along the LOS, and the bisector ranges between the line core (0\%) and the outer wings (100\%). The wavelength position of each bisector at a given percentage can be computed using linear interpolation between both sides of the spectral line. The Doppler shifts measured from line bisectors at a given intensity level contain information from a significant fraction of the atmosphere, and therefore it is difficult to ascribe them to any particular layer. This limitation of bisector analyses has long been recognised \citep{Maltby1964, Rimmele1995b}.

Typically, the bisectors trace a ``C'' shape, which results from spatially averaging different types of spectra over granulation \citep{dravins1981,asplund2000}. Bright, asymmetric, and blueshifted spectra from granules are averaged together with dark, asymmetric, and redshifted spectra from the intergranular lanes. Since more area is covered by granules, both groups of spectra are unbalanced. Furthermore, bisector reversals (inverted `C' shape) also appear in the literature \citep[e.g. ][]{Wiehr1984, Rimmele1995b}. 
\citet{WestendorpPlaza2001} associated those reversals to elevated flow channels. 

A comparison of LOS velocities inferred from the bisector method and the spectral-line inversion code SIR \citep{RuizCobo1992} was made in the visible spectral range. \citet{BellotRubio2006} used the strong photospheric \ion{Fe}{i} 5576~\AA\ line with zero Land\'e factor $g$$=$$0$ for that study. The authors obtained good agreements comparing both methods, in both high and deep layers. The LOS velocities of bisectors computed near the continuum level (60\%--80\%) are very well correlated with LOS velocities at optical depths of $\log \tau_{5000} = -0.6$, whereas bisectors between 2 and 20\% of the line are correlated well with optical depths of $\log \tau_{5000} = -1.8$.

The possibility of inferring the height dependence of the LOS velocity is especially useful when we observe spectral lines that form in a broader range of heights; for example the Fraunhofer \ion{Na}{i}, \ion{K}{i}, and \ion{Ba}{ii} $D$ lines, or the \ion{Si}{i} transition at 10827~\AA. The latter covers a narrower range of heights in comparison with the other mentioned candidates, up to the upper photosphere \citep[e.g.][]{Bard2008,Shi2008,Sukhorukov2012}. Furthermore, the line is typically observed together with the \ion{He}{i} 10830~\AA\ triplet, as it serves as an auxiliary line to constrain the atmospheric parameters in the photosphere. Numerous studies have examined these lines together in arch filament systems \citep[e.g.][]{Lagg2007,Yadav2019,GonzalezManrique2020}, filaments \citep[e.g.][]{kuckein12a,Xu12}, Ellerman bombs \citep[e.g.][]{Rezaei15}, sunspots and pores \citep[e.g.][]{Xu2010,Joshi14}, flares \citep{judge14,kuckein15}, 
or independently, for example, in small-scale magnetic bright points \citep{kuckein19} or sunspots \citep[e.g.][]{Rezaei12b,Felipe16,orozco17}.
This unique spectral region will also often be scanned by DKIST and EST. Therefore, we focus our efforts on developing a method to compute bisectors and testing it on the mentioned \ion{Si}{i} line. Our aim is to provide a tool  to estimate the vertical stratification of the LOS velocity that is simple, fast, and reliable. We hope for this tool to be integrated as an on-site quick-look method that provides an estimation of the ongoing observations running on DKIST and EST, and can extract information from the ensuing reduced data products. Moreover, this tool can be used to define better initial guess atmospheres for the mentioned non-local thermodynamic equilibrium (NLTE) inversion codes, which is an objective shared by recent works like \citet{Gafeira2020}.	

%
%

\section{Methodology}\label{SEC2}


\subsection{Synthesis of spectral profiles}\label{SEC2.1}

\begin{figure}
\begin{center} 
 \includegraphics[trim=0 0 0 0,width=9.0cm]{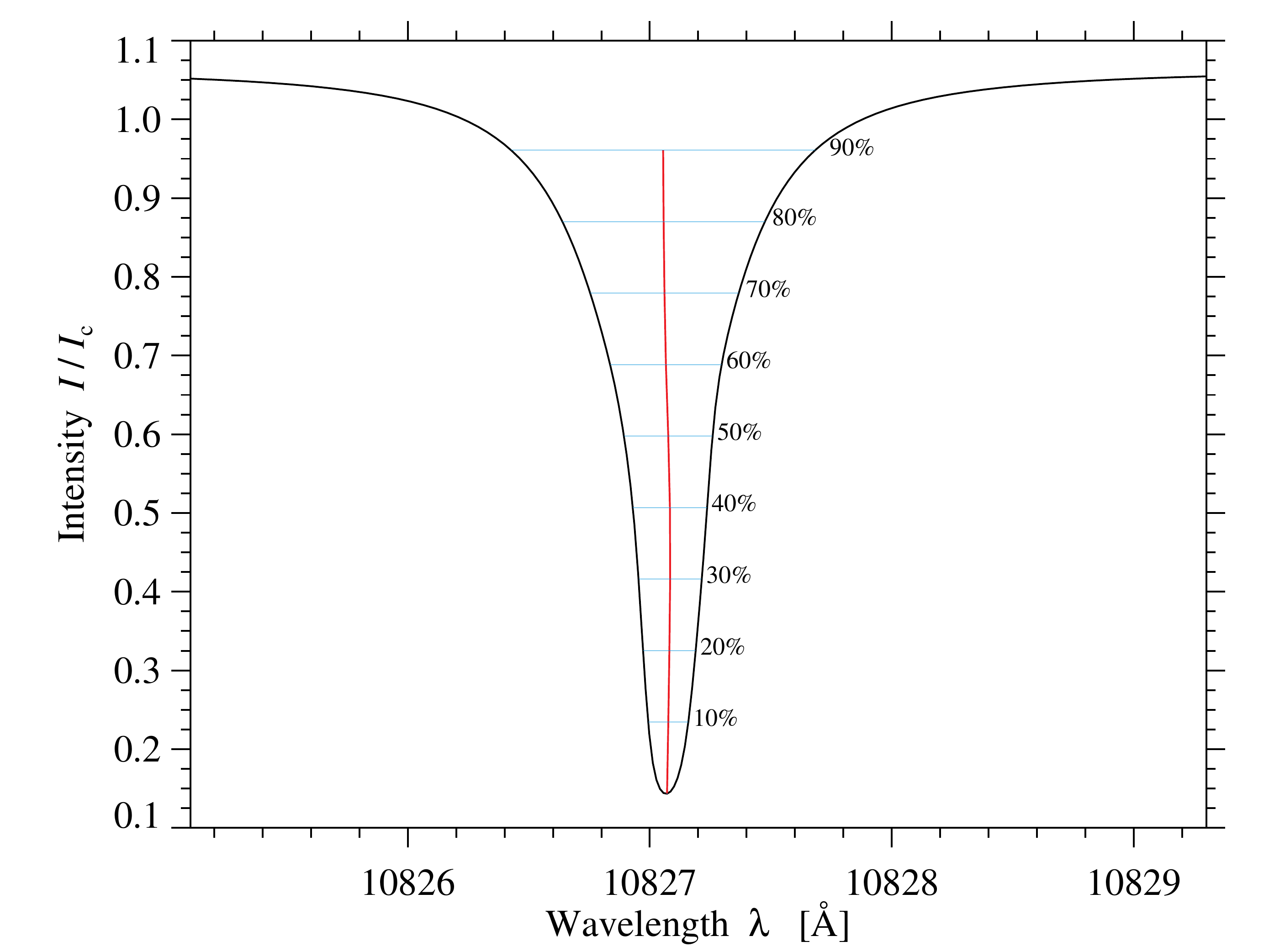}
 \vspace{-0.4cm}
 \caption{Reference intensity values used for computing the bisectors of the \ion{Si}{i} 10827 \AA\ line. The background profile corresponds to the spatially averaged intensity over the entire FOV presented in Figure~\ref{context}, normalised to the local continuum intensity $I_c$.}
 \label{FIG_BISEC}
 \end{center}
\end{figure}

We make use of the RH code \citep{Uitenbroek2001} to synthesise the intensity profiles. This code can compute the atomic populations of the levels involved in the transitions under NLTE conditions. We use two types of atmospheric models: a 1D semi-empirical atmosphere and 3D realistic numerical simulations. In the first case, we employ the FALC atmosphere \citep{Fontenla1993} for computing the response functions (RFs) \citep[e.g.][]{Landi1977} of the spectral line to perturbations of the atmospheric parameters. In the second case, we use the snapshot 385 of the Enhanced Network simulation presented in \cite{Carlsson2016} and developed with the Bifrost code \citep{Gudiksen2011}.  The latter reproduces a bipolar network patch surrounded by quiet Sun areas (Figure~\ref{context}), and has been used in multiple works \citep[among others][]{QuinteroNoda2016,Stepan2016,Golding2016,Sukhorukov2017,daSilvaSantos2018}; see also the review of \cite{Carlsson2019}. It is publicly available\footnote{\url{http://sdc.uio.no/search/simulations}} and  therefore increases the repeatability of our work.

We assume disc centre observations, that is $\mu=1$, where $\mu=\cos(\theta)$ and $\theta$ is the heliocentric angle. We do not include any spatial degradation in our studies, that is, we use the original horizontal pixel size of 48~km, and \cite{Asplund2009} gives the abundance of the different atomic species. Previous works noted that there is a lack of small-scale random motions in this simulation \citep{Leenaarts2009} leading to narrower profiles than those detected in solar observations \citep[see also][]{delaCruzRodriguez2012}. For this reason, we add a microturbulence of 1.0~km/s constant with height. This value is estimated from the microturbulence stratification of the FAL models at around 500~km \citep[e.g.,][]{Fontenla1990}. We use a spectral sampling of 10~m\AA \, and no spectral degradation is considered in this work.

We compute the atomic populations for the silicon transition under NLTE using the simplified atomic model presented in \cite{Bard2008}. The model contains 23 levels and 149 line transitions, and allows the computation of the \ion{Si}{i} 10827~\AA \ spectral line in reasonable times, that is, around 1 CPU minute per pixel. Moreover, as we use the parallel version of RH presented in \citet{Gafeira2020}, the computational time for the entire map is affordable: around 30 CPU hours in a single server with 48 physical cores. However, we highlight the fact that for more computationally expensive studies, such as for example NLTE inversions, a more simplified atomic model
should be used, like the one presented by \cite{Shchukina2017}.


\subsection{Estimation of LOS velocities from the \ion{Si}{i} line}\label{SEC2_2}

We use two methods to infer the LOS velocities. On the one hand, we employ a bisector analysis for line profile intensities between 10 and 90\% of the spectral line depth. The bisectors were obtained in 10\%~intensity steps using an intensity linear interpolation for each line wing. The wavelength positions were then calculated, computing the centre position between both sides of the line for each bisector level (Fig.~\ref{FIG_BISEC}). On the other hand, for the line core intensity, we use a second-order line core fit to derive the Doppler shift. Both methods were used for the entire field of view (FOV) presented in Figure~\ref{context}.

\section{Results}\label{SEC3}


\subsection{Response functions}\label{SEC3.1.1}



\begin{figure}
\begin{center} 
 \includegraphics[trim=0 0 0 0,width=8.3cm]{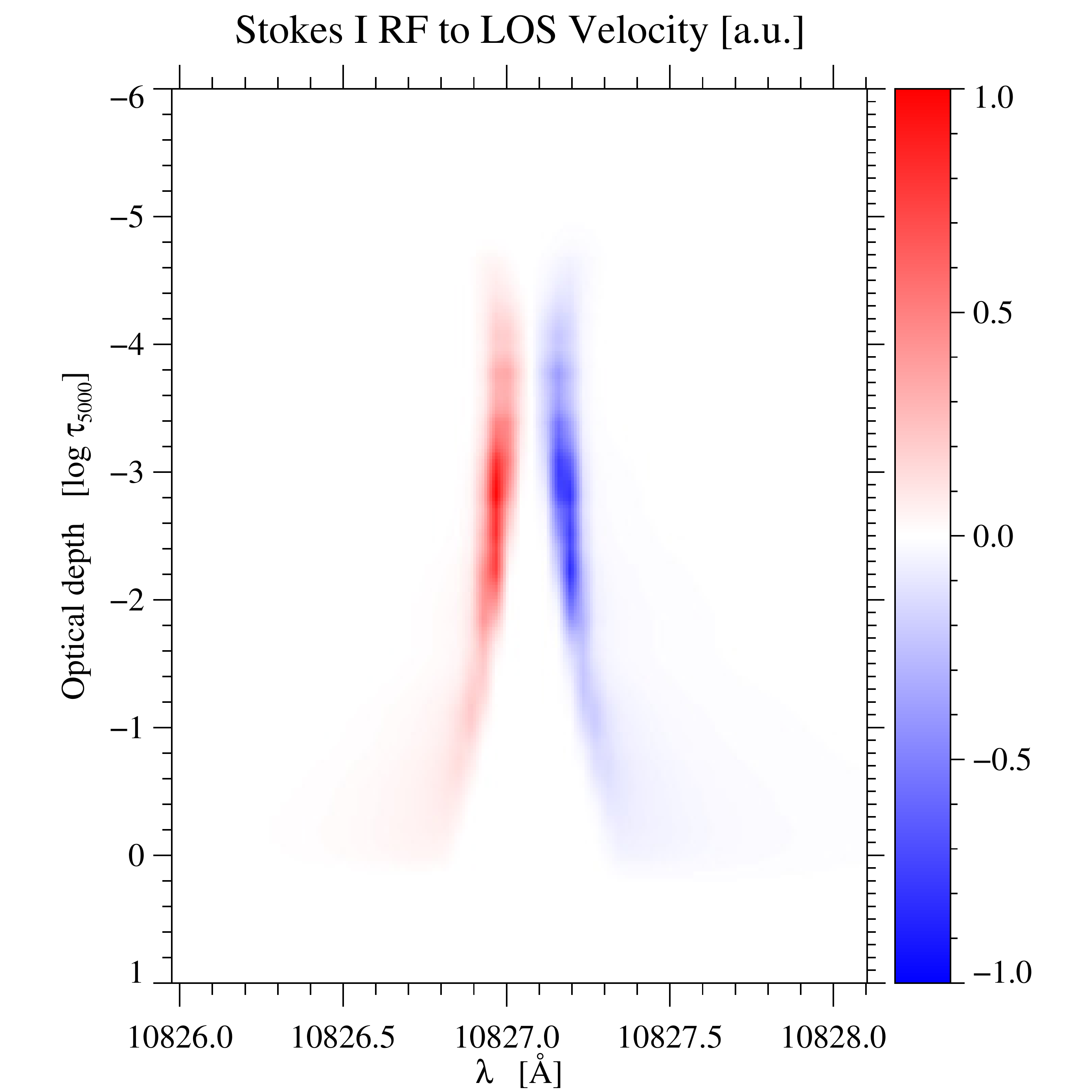}
 \vspace{-0.1cm}
 \caption{Two-dimensional plot of the NLTE Stokes $I$ RF to changes in the LOS velocity starting from the FALC model. White areas designate regions of no sensitivity to changes in the atmospheric parameters, while red and blue colours indicate opposite signs of the RF.}
 \label{2DRF}
 \end{center}
\end{figure}

We first compute the numerical RF to perturbations on the LOS velocity for the intensity profile of \ion{Si}{i} 10827~\AA. We start from the original FALC model with null LOS velocity and we modify it with a perturbation of $\delta x=0.1$~km\,s$^{-1}$ at each point in optical depth. The results, included in Figure~\ref{2DRF}, show that the spectral line covers a wide range of optical depths, wider than that found for traditional photospheric lines \citep[e.g.][]{CabreraSolana2005}. In addition, there is a smooth transition between the wings that is sensitive to lower layers, and line-core wavelengths reaching up to $\log \tau \sim -4.3$.

\begin{figure}
\begin{center} 
 \includegraphics[trim=0 0 0 0,width=8.3cm]{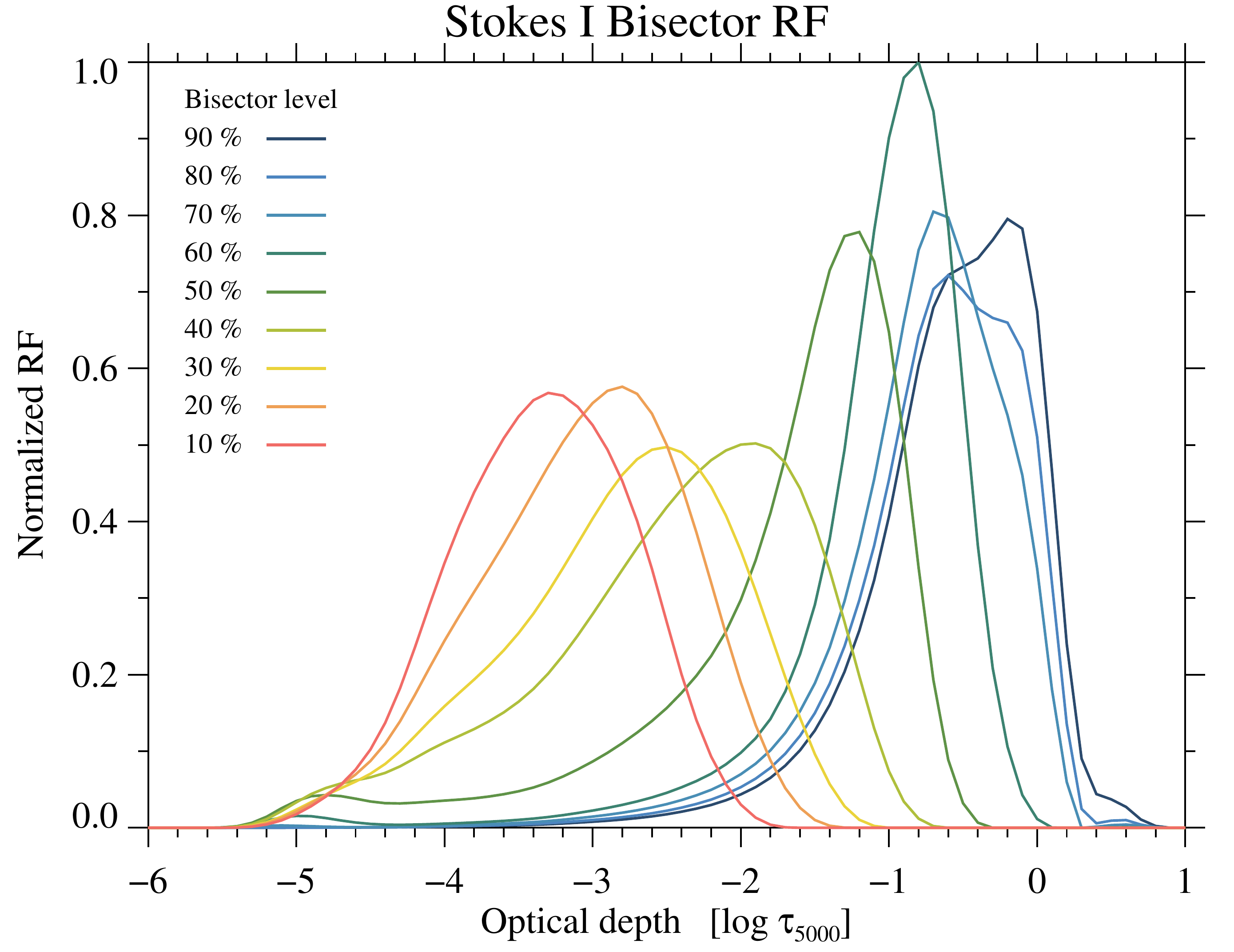}
 \vspace{-0.1cm}
 \caption{Response functions to the bisector method. Each colour represents different spectral line intensity levels where the bisectors are computed.}
 \label{RFbisec}
 \end{center}
\end{figure}

\begin{figure*}
\begin{center} 
 \includegraphics[trim=0 0 0 0,width=18.0cm]{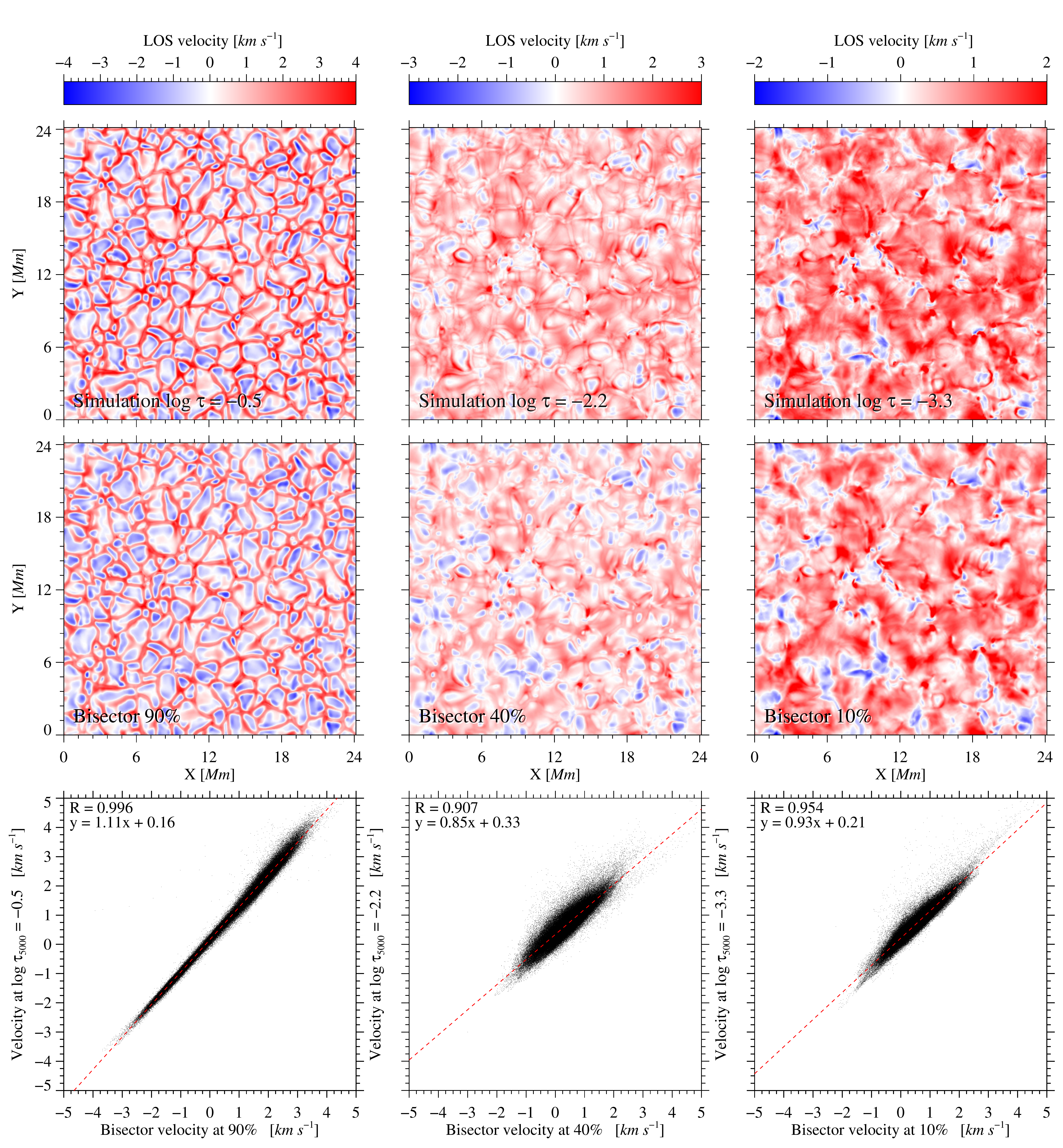}
 \vspace{-0.2cm}
 \caption{Line-of-sight velocity maps from the original simulation (top) and those inferred using the bisector method (middle). We include in the bottom row a scatter plot and the Pearson correlation between both maps. Each column designates a different height (in the case of the simulation) and bisector velocity at different line depths.}  
 \label{Simcompare}
 \end{center}
\end{figure*}

\begin{figure*}
\includegraphics[trim=0 0 0 0,width=18.0cm]{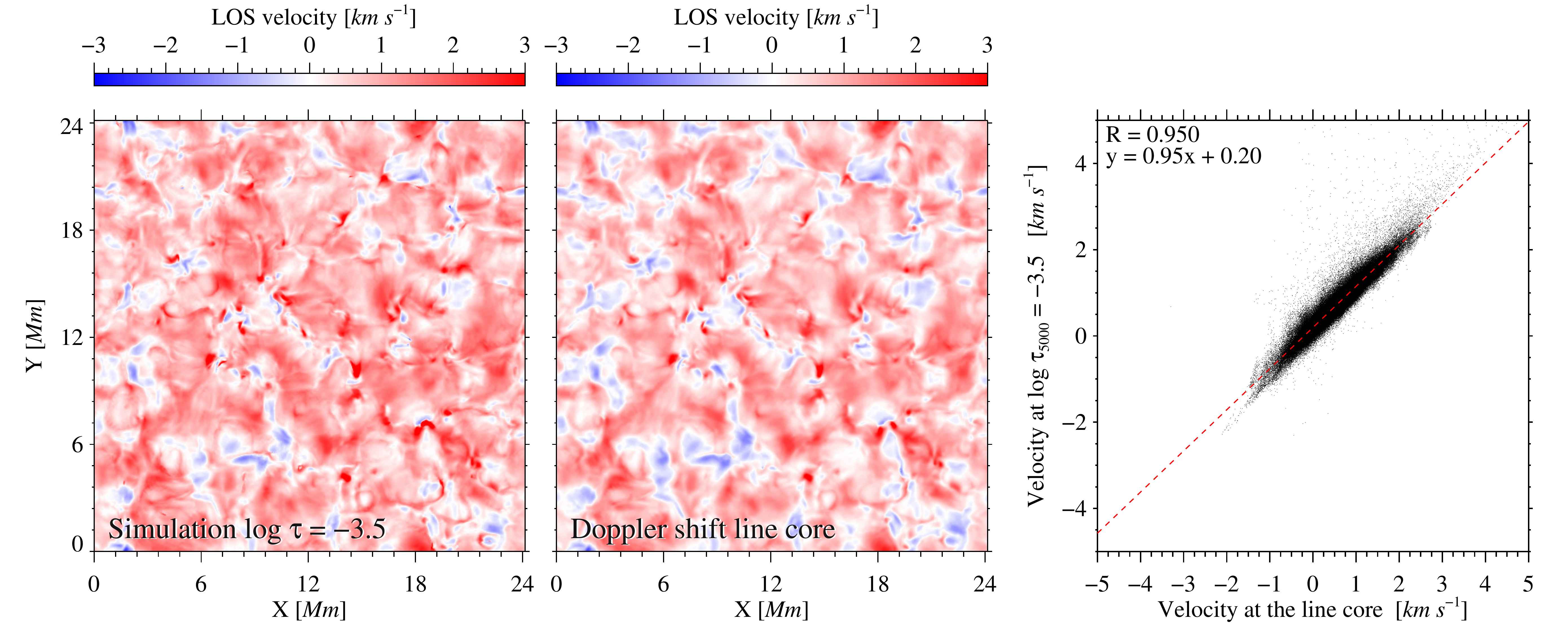}
\caption{Comparison between the LOS velocity from the simulation (left) and that obtained from a line core fit of the spectral line (middle). The leftmost panel includes a scatter plot and the Pearson correlation between both maps.}
\label{correlationcore}
\end{figure*}

In order to derive the sensitivity of the bisector at different depths, a numerical transformation is applied to the depth-dependent LOS velocity RFs. This numerical transformation can be interpreted as the change of the abscissa of each bisector level when a LOS perturbation is introduced at each optical depth. We use the theory introduced in \cite{RuizCobo1994}, which allows to obtain the RF to any parameter that has been obtained from the spectrum as a very simple operator acting over the LOS velocity RF of the intensity. 

In our case, the bisector $\lambda_\mathrm{bis}$ is obtained after a linear
interpolation of the intensity values $I_i\equiv I(\lambda_i)$ around bisector
levels $I_{\mathrm{bis},j}$. The RF of each bisector position to the LOS velocity
perturbations $R^{\mathrm{bis},j}_\mathrm{LOS}$ is then obtained after
applying the chain rule: first, we compute the derivative of
the bisector with respect to the intensity at the wavelengths that
we are using to interpolate $(\lambda_i)$; then we multiply these derivatives by the RF
at each wavelength ($R_\mathrm{LOS}(\lambda_i))$; and finally, all the products are summed up.
After some simple algebra, we obtain:
\begin{equation}
    R^{\mathrm{bis},j}_\mathrm{LOS}= \sum_{i=1}^4 c_i R_\mathrm{LOS}(\lambda_i)~,
\end{equation}
where the coefficients $c_i$ are given by: 
\begin{equation}
c_i=\frac{(\lambda_i-\lambda_k)*(I_k-I_{\mathrm{bis},j})}{2(I_i-I_k)^2} 
,\end{equation}
where $k=i+(-1)^{(i+1)}$.

The results for the RF to the bisector are included in Figure~\ref{RFbisec}. Each percentage level has a distinct peak and a certain width that covers a range of atmospheric layers. The latter property indicates that they cannot be associated to a single atmospheric layer. However, at the same time, each bisector level has a peak at a specific optical depth that linearly grows from lower layers (at $\log \tau_{5000} \sim -0.2$ for the bisectors at 90\% intensity) to higher atmospheric heights up to $\log \tau_{5000} \sim -3.5$ (bisectors taken at 10\%), indicating that we are examining different atmospheric layers when computing the bisectors of the \ion{Si}{i} 10827~\AA \ transition at different intensity levels. These results are compatible with the Stokes $I$ RF to the LOS velocity presented in Figure~{\ref{2DRF}}.


\subsection{Bisector results}\label{SEC3.2}

We compute the bisector velocity maps for the entire FOV presented in Figure~\ref{context}, and compare them with the original simulated atmosphere using Pearson's linear correlation coefficient. To do so, we transform the input atmosphere from geometrical height to optical depth at continuum wavelength at 5000~\AA. We find good correlation values in the range of 90-99\% between the inferred LOS velocities and the input atmosphere at selected optical depths. As an example,  Fig.~\ref{Simcompare} presents the original atmosphere (top), the bisector inferred velocities (middle), and their correlation (bottom) for three selected bisector line depths. In the case of the far-wing wavelengths (bisector at 90\%), the spatial distribution of LOS velocities is very similar to that from the deep photosphere with a correlation of 99.6\%. In the second case, when computing the bisectors at 40\%, the differences are noticeable with more blueshifted areas in the inferred velocities and with a lower correlation than before (90.7\%). Moving to the last selected case, where bisectors are taken close to the line core at only 10\% line depth, we find good results again. The spatial distribution is very similar (although with some blueshifted areas in the inferred velocities that are not present in the original simulation), and the correlation is above 95\%.

\subsection{Line-core Doppler shifts}

The bisector technique is based on the difference of intensity values taken at both sides of the central wavelength of the transition. Therefore, we cannot reach the deepest line intensity, that is, the line core, with this method. However, we want to optimise the use of this synthetic data. We do this by testing the accuracy of a traditional line-core fit for estimating the LOS velocity through line-core wavelength shifts. We follow the same concept whereby we use a simple and fast approach, aiming to use it for quick analysis of the large data maps generated by future missions. We employ the internal IDL routine \textit{polyfit.pro}, and fit the intensity spectra with a second-order degree for the entire FOV in less than 10 seconds. The spectral range used to fit the core was the
minimum intensity $\pm$ 50~m\AA. The results are plotted in Figure~\ref{correlationcore}. We can see that the spatial distribution of the LOS velocity at  $\log \tau_{5000}=-3.5$ (leftmost panel) resembles that obtained from the line core fit (middle column). As before, there are some areas that show blueshifted velocities that are not present in the original snapshot but the general inferred velocity pattern is similar, which is also reflected in the Pearson correlation of 95\% (rightmost panel).

\subsection{Height coverage}
\begin{figure}
\includegraphics[width=\columnwidth]{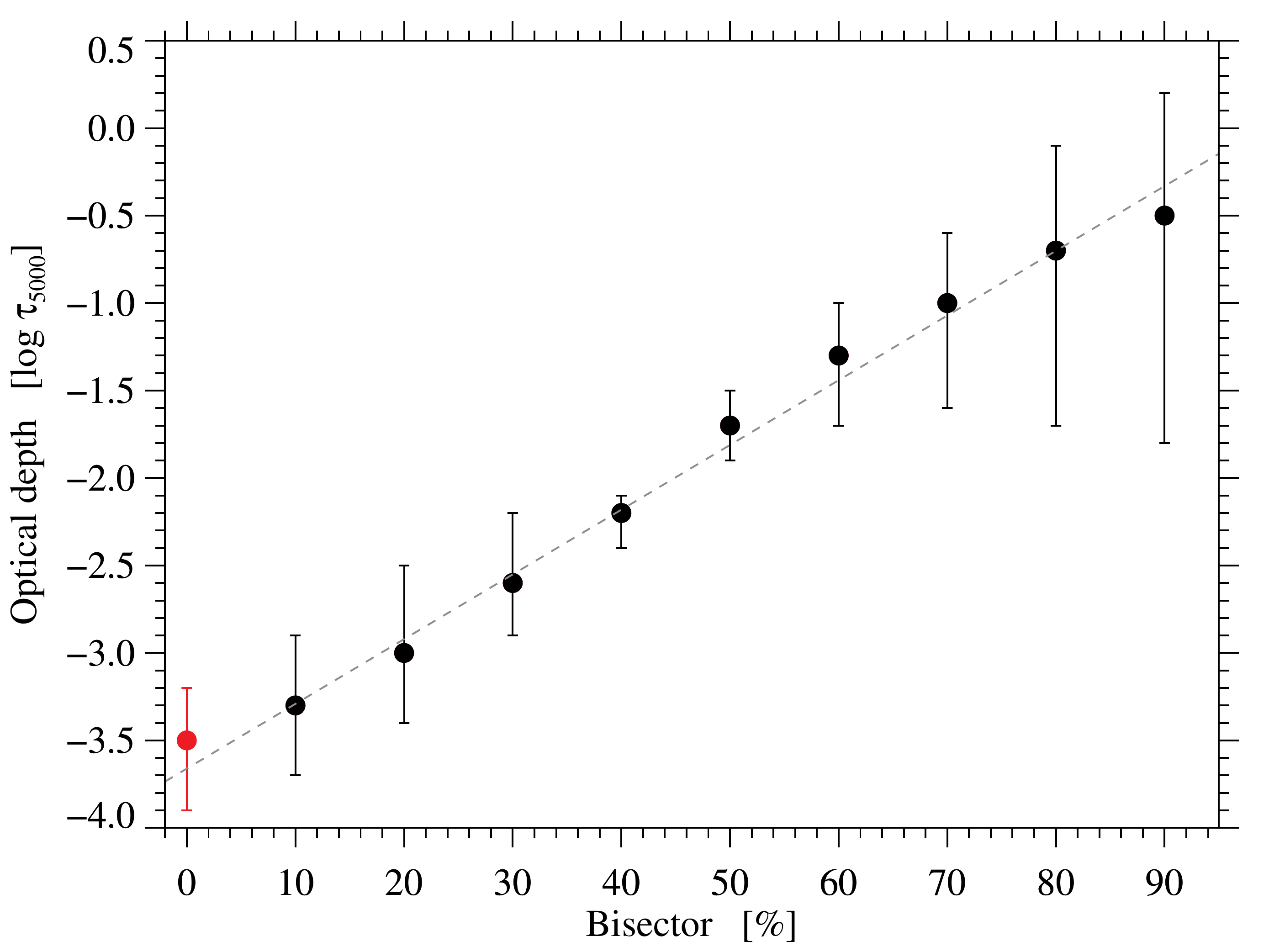}
\caption{Comparison between \ion{Si}{i} 10827\,\AA\ LOS velocity maps of the height-dependent bisectors method and LOS maps of the simulation snapshot transformed to optical depth. The filled circles depict the highest correlation for each bisector map at a given line depth. The dashed line illustrates the linear fitting between all the bisectors (black dots). The red filled circle designates the results from the line-core fit using a second-order polynomial. Uncertainty bars indicate the optical depths where the correlation is higher than 90\% for each bisector level.}
\label{correlation}
\end{figure}

In the previous sections we demonstrate the robustness of the bisectors method and the association of these bisectors to different optical depths in the quiet Sun for the \ion{Si}{i} line. Now, we aim to find a relation between the bisector percentages and optical depths in order to establish a linear relation between both quantities. Therefore,
we complement the previous results with a statistical study computed for the entire simulation snapshot. The results of this comparison are shown in Figure~\ref{correlation}. Black filled circles indicate the highest correlation between the bisector percentage and the optical depth from the simulations. We computed the correlation of each bisector LOS velocity map with the original simulation LOS velocity map at different optical depths. There is a linear relation between the bisector percentages and the optical depths that results in:
\begin{equation} \label{eq1}
\log \tau = 0.037x - 3.66,
\end{equation}
where $x$ is the bisector intensity level as a percentage. We note that although we did not consider the line core in this computation, we include the results from the line-core fit (highlighted as a red dot) in the same plot. The results are in agreement with the RF studies (see Figs.~\ref{2DRF} and \ref{FIG_BISEC}), where the bisector velocities obtained deeper in the line correspond unequivocally to higher layers in the atmosphere. In addition, we can cover the velocity stratification up to $\log \tau_{5000}\sim-3.5$ by combining the bisector method and the line-core fit. 

%
%

\section{Summary}\label{SEC4}

We aimed to examine the capabilities of fast methods to infer the LOS velocities from future observations of the photospheric \ion{Si}{i} 10827~\AA\ line. In particular, we studied the accuracy of bisector analysis and a line-core fit of the line-core Doppler shift. We used synthetic profiles obtained from the Bifrost enhanced network simulation solving the radiative transfer equation with the RH code. On this occasion, we employed an atomic model of \ion{Si}{i} to compute the atomic populations of the different levels assuming NLTE effects. 

We first analysed the intensity RFs to changes on the LOS velocity using the FALC semi-empirical atmosphere. We found that the \ion{Si}{i} transition is sensitive from the deeper photosphere up to $\log \tau=-3.5$ in the line core. Moreover, there is a smooth transition from the far wings (sensitive to deeper layers) to the line core (sensitive at higher layers) with no degeneracy. We added the RF to the bisectors finding that, although each bisector level covers a specific range of optical depths, the location of their highest value increases linearly with height.

We then employed the NLTE synthetic profiles obtained from the 3D Bifrost enhanced network simulation to estimate the accuracy of the bisector method. Comparison with the original simulation indicates that the results from the bisector method are reliable for a broad range of wavelengths. The spatial distribution of the inferred velocities shows, in general, a good correlation with the original simulation velocities, although there is a higher incidence of blueshifted areas in the inferred velocities. Nevertheless, the obtained velocities are similar to those from the simulation, which is verified through very high (larger than 90\%) Pearson correlation values. We also added a line-core fit, only
for the line-core intensity, to determine the LOS velocities at those wavelengths. The results again resemble those from the original simulation at higher layers in the atmosphere. Finally, when computing the optical depths where the correlation of different bisector line depths is highest, we obtained a linear relation (\ref{eq1}). This relation, obtained in this study for the quiet Sun, means that there is no degeneracy when computing bisectors, that is, wavelengths closer to the line core correspond to higher atmospheric layers in the atmosphere.

The tools presented in this study have various applications. For instance, we can use them as a quick estimator of LOS velocities for real-time tracking of the observations; for example, the bisector analysis for the complete simulation snapshot took less than 20 CPU seconds. Moreover, we can use them to set initial LOS velocities in a guess atmosphere for future NLTE inversions of \ion{Si}{i} 10827\,\AA\ to be performed by SNAPI, STiC, or NICOLE,  for example. The latter approach may allow a reduction of the number of inversion iterations or degrees of freedom for the LOS velocity, thus reducing the computation time of complex NLTE inversion codes. Therefore, albeit simple, this technique seems to be robust and reliable for the \ion{Si}{i} 10827~\AA \ spectral line, and therefore we recommend its use for the applications mentioned above on future DKIST and EST data.

We find a similar technique in the literature for inferring the LOS velocity stratification, that is, the $\lambda$-metre method \citep{Stebbins1987}, which has been used in several studies \citep[e.g.][and references therein]{Carlsson1998,Kostik2007, Shchukina2009}. We plan to expand these results comparing the bisector analysis with the $\lambda$-metre method to estimate which one is more reliable for different solar conditions. Finally, we emphasise that the Bifrost enhanced network simulation was chosen because the complete run is publicly available, assuring the reproducibility of our work. However, this means that we can only guarantee the accuracy of the methods tested in this publication for quiet Sun observations. Therefore, we should expand these results in the future with more complex scenarios such as those presented by \cite{Hansteen2017,Hansteen2019} or in stronger magnetic field concentrations like that of \cite{Cheung2019}.

%
%

\begin{acknowledgements}
SJGM acknowledges the support of the project VEGA 2/0048/20 and the support by the Erasmus+ programme of the European Union under grant number 2017-1-CZ01-KA203-035562 during his 2019 stay at the Instituto de Astrof{\'i}sica de Canarias. SJGM is grateful for the financial support received from the EST Science Meeting in Sicily 2018 to attend the meeting and present this work. This work was supported by the Research Council of Norway through its Centres of Excellence scheme, project number 262622. This work has been partially funded by the Spanish Ministry of Economy and Competitiveness through the Project RTI-2018-096886-B-C53. This research has made use of NASA’s Astrophysics Data System. Funding from the Horizon 2020 projects SOLARNET (No 824135) and ESCAPE (No 824064) is greatly acknowledged.
\end{acknowledgements}


\bibliographystyle{aa}
\bibliography{aa-jour,sergio}

\end{document}